\documentclass[12pt]{iopart}

\usepackage{graphicx}
\usepackage{color}
\usepackage{dcolumn}
\usepackage{bm}

\begin{document}
\title[]{Quantum Criticality and Population Trapping of
  Fermions by Non-Equilibrium Lattice Modulations}

\author{Regine Frank$^{1,2}$}

\address{
$^{1}$ Institut f\"ur Theoretische Festk\"orperphysik Wolfgang-G\"ade-Strasse 1,\\
{\color{white}$^{1}$} Karlsruhe Institute of Technology (KIT), 76131 Karlsruhe, Germany,\\
$^{2}$ Institut f\"ur Theoretische Physik, Optics and Photonics,\\
{\color{white}$^{2}$} Center for Collective Quantum Phenomena (CQ) and\\
{\color{white}$^{2}$} Center for Light-Matter Interaction, Sensors \& Analytics (LISA+),\\
{\color{white}$^{2}$} Eberhard-Karls-Universit\"at T\"ubingen, Auf der Morgenstelle 14, 72076 T\"ubingen}
\ead{r.frank@uni-tuebingen.de}
\date{\today}

\begin{abstract}
An ultracold gas of interacting fermionic atoms in a three-dimensional optical 
lattice is considered, where the lattice potential strength is periodically modulated. 
This non-equilibrium system is non-perturbatively described by means of a
Keldysh-Floquet-Green's function approach for Mott-Hubbard systems employing a generalized dynamical
mean field theory (DMFT). Strong repulsive interactions between different
atoms lead to a Mott insulator state for the equilibrium system, but the
additional external driving at zero temperature yields a non-equilibrium
quantum critical behavior, where an infinite number of
Floquet states arise and a transition to the liquid and conducting phase is given.
\end{abstract}

\maketitle

\section{Introduction}

The melting of crystals and ice has been intriguing ever since Albert Einstein
proposed his quantum theory of solids \cite{Coleman}. Novel results in quantum
and non-equilibrium physics are pushing us to improve our understanding of how
quantum matter behaves at ultra-low temperatures. In condensed matter physics
quantum dynamics can only be analyzed under the premise of severe influences
due to strong coupling to the environment. This aspect makes it challenging
to prepare  and control  quantum states far from equilibrium. 
Thus ultracold gases are perfect systems to study the pure influences
of quantum non-equilibrium effects in first instance, since the properties 
of such systems  can be tuned almost without restrictions. At this point 
the reader might ask the question ``Why {\em non}-equilibrium~?''. 
Whereas calculations in thermodynamically equilibrated systems are well 
established in theory, and rather complicated structures requiring high-end 
numerics can be solved, it is still a challenge to determine results for driven
systems \cite{Trotzky}. Nevertheless those systems are experimentally and
technologically interesting \cite{Daley, Giamarchi1, Oka1Oka2, Schneider, Zwerger}. 
Even if a system is in the steady-state regime, it still does not
reside in thermodynamical equilibrium if it is driven, {\em e.g.}, by a nonzero
current or a difference in the potential \cite{Heidrich}, and the 
reflection of that fact in theory is highly desirable.\\

\section{Theoretical Framework}

Recently, intriguing experiments 
on non-equilibrium dynamics of ultracold gases, both fermionic as well as bosonic,
 \cite{Greiner, Sherson, Bakr, G-2005, Weidenmueller2010, Weidenmueller2009} have become
 possible. Among many aspects, also the long time limit of such
 non-equilibrium systems has been studied \cite{Kinoshita, Giamarchi, Esslinger2010}, 
 and applications using non-equilibrium processes as 
 a tool for transport in so called quantum
 ratchets have been investigated \cite{Weitz}. 

The Hubbard Hamiltonian is one of the most relevant models for investigating
strongly correlated systems in condensed matter theory, of both bosonic as
well as  fermionic nature \cite{Rigol, Jaksch, Hofstetter, Pozashennikova, Frank}.
In this article we study the characteristics of an interacting ultracold
Fermi gas exposed to periodic modulations of the
optical lattice strength.
This configuration corresponds to a stationary non-equilibrium
state, which requires suitable techniques such as the Keldysh formalism. 
The periodic modulation leads to a 'dressing' of atoms, well known from the
application of light fields in quantum optics. We discuss a 
dynamical mean field theory (DMFT) solution \cite{Freericks} for the
Floquet-Keldysh \cite{Haenggi} approach. The non-equilibrium-caused dressed
states arise as Floquet side bands.
In the Mott-Hubbard gap we derive a complicated modulation-induced structure
of many particle states, and therefore a transition from the Mott insulting
regime to a liquid phase
which leads to a  finite conductivity. The occupation number for these gap
states is investigated and we find a trapping of population which results in
an 'inversion' for increasing modulation frequencies.

\vspace*{1.5cm}
\begin{figure}
\qquad\qquad\qquad%
\begin{center} \scalebox{0.5}[0.5]{\includegraphics[clip]{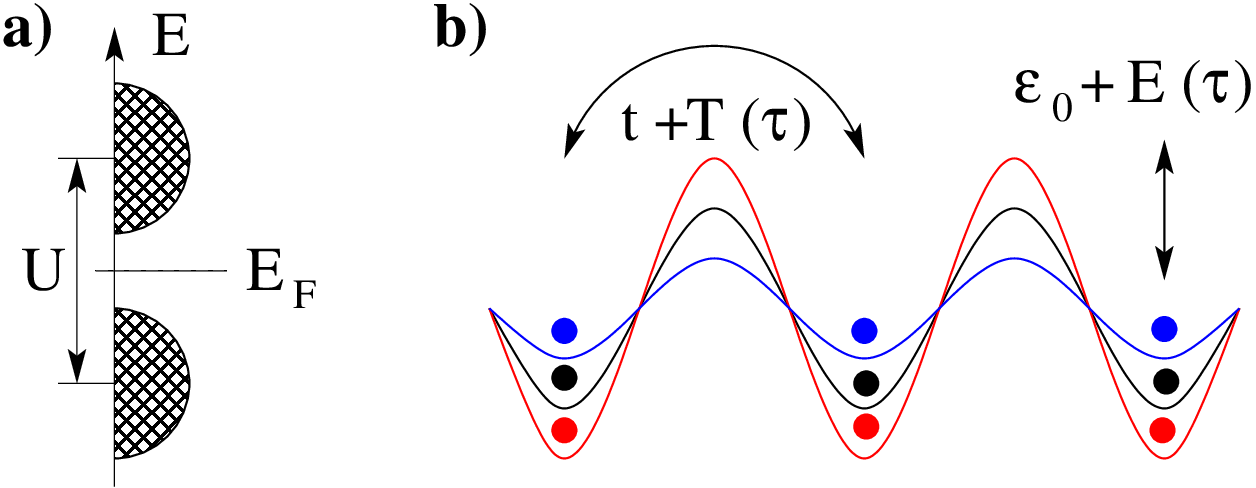}}\end{center}
\caption{(Color online)
(a) The strong local interaction $U$ between atoms in the
equilibrium system causes the single energy band to split into the lower and the upper
Hubbard band, separated by an excitation gap. 
(b) Ultracold fermionic atoms in a 3-dimensional  optical lattice (black). 
The optical lattice is
periodically modulated with the frequency $\Omega_L$, yielding a minimum (red) and a
maximum (blue) potential modulation. The periodic modulation introduces an additional,
time-dependent contribution to the local atom energy $E(\tau)$ and the
atomic  tunneling amplitude $T(\tau)$ (see text). 
The  sketched potentials of this non-equilibrium system therefore represent
three different snapshots in progress of the external periodic modulation.
}
\label{scheme}
\end{figure}
The Fermi gas in the modulated optical potential is schematically shown in
Fig. \ref{scheme}. The Mott insulator state in equilibrium is
characterized by a tight binding model with strong repulsive onsite interactions
$U$ experimentally determined by a Feshbach resonance. 
The onsite interaction leads to a band splitting and the establishment of the
characteristic  Mott-Hubbard gap. 
Periodic modulations of the optical lattice potential $V(\tau)$ influence the
behavior of the tunneling  $t$  from one lattice site to the nearest neighbor as well as the
onsite energies $\epsilon_0$ at each lattice site.
The equilibrium tunneling amplitude $t$ must be replaced by $t+T(\tau)$; 
the equilibrium onsite energy has to be replaced by
$\epsilon_0 + E(\tau)$ as well.  These time $\tau$ dependent terms are  
periodic themselves. We consider the following Fermi-Hubbard Hamiltonian
\begin{eqnarray}
\label{Hamilton1}
H (\tau) &=&   
 \sum_{i, \sigma}                     
\left[ \epsilon_0 + E(\tau) \right]  
c^{\dagger}_{i,\sigma}c^{{\color{white}\dagger}}_{i,\sigma} \\
&-&\!\!\!\sum_{\langle ij \rangle, \sigma}    
\left[ t  +         T(\tau) \right]  
c^{\dagger}_{i,\sigma}c^{{\color{white}\dagger}}_{j,\sigma} 
 +  
\frac U 2 \! \sum_{i,\sigma} \!\!  
c^{\dagger}_{i,\sigma}c^{{\color{white}\dagger}}_{i,\sigma} 
c^{\dagger}_{i,-\sigma}c^{{\color{white}\dagger}}_{i,-\sigma}. 
\nonumber  
\end{eqnarray}
The onsite repulsion $U$ is considered to be not majorly affected 
by temporal changes,  because it is large  
compared to possible effects  due to lattice oscillations. Therefore, 
$U$ is taken as constant in the following.
The index $i$ labels the lattice site and $\sigma$ the spin; 
$\langle ij \rangle$ implies summation over nearest neighbors;
$ c^{\dagger}_{i,\sigma}$ and $c^{{\color{white}\dagger}}_{i,\sigma}$ create
(annihilate) a fermionic atom with spin $\sigma$ at lattice site $i$.
The time-dependent contributions due to the periodic modulation of the
potential are given by 
\begin{eqnarray}
\label{E_und_T_von_tau}
 E(\tau) =  E\cos(\Omega_L \tau)  \qquad\quad  T(\tau) =  T\cos(\Omega_L \tau), 
\end{eqnarray}
where $\Omega_L$ is the frequency of the lattice modulation, $\tau$ is the
system time and $E$ and $T$ are the respective amplitudes of the energy and
the hopping or
tunneling  contribution. Note, during the numerical evaluation the parameters $t$ 
and $T$ have to be chosen such that for any time $\tau$ the kinetic term 
in the Hamiltonian, Eq. (\ref{Hamilton1}), does not change sign, i.e. 
$T(\tau) < t \quad \forall\quad \tau$.
The maximum of the hopping amplitude $t$ is set to be equal $8\,D$, where $D$ is the half bandwidth.

Driven systems, such as fermionic atoms in a modulated lattice potential,
experience an energy exchange with their exterior and therefore do not reside
in a state of thermodynamical equilibrium.
Due to the non-equilibrium character the system response, as e.g. expressed in
the Green's function, depends on two distinct time arguments. For instance on a
defined starting point and the elapsed time, or after a change of the reference
frame on the center-of-mass time and the relative time coordinate. The evolution of an 
equilibrium state in contrast is usually sufficiently described by the 
relative time alone. To account for this, Schwinger \cite{Schwinger} and in his
footsteps Keldysh \cite{Keldysh} designed a theoretical framework. 
The system in the distant past ($\tau =-\infty$) is considered in a defined
state, the interaction is then slowly switched on as time progresses, 
the system evolves to the
present where measurements are considered, and then it evolves via
($\tau =+\infty$) back to  ($\tau =-\infty$). Along this path the interaction is
switched off. This particular time contour is also called Schwinger-Keldysh
contour. The time arguments of the Green's function may be found on
upper branch of the contour, evolving from $\tau =-\infty$ to $\tau =+\infty$ or
the time argument may reside on the lower branch,  from $\tau =+\infty$ to
$\tau =-\infty$. A matrix Green's function is considered according to
\begin{eqnarray}
\tilde{ {\bf G} }   (\tau_1,\tau_2) = 
\left(
\begin{array}{lcl}
G^{++}(\tau_1,\tau_2)   & \qquad &  G^{+-}(\tau_1,\tau_2)   \\
G^{-+}(\tau_1,\tau_2)   & \qquad &  G^{--}(\tau_1,\tau_2)
\end{array}
\right)
\end{eqnarray} 
where the superscripts denote on which branch of the contour ($+=$ upper; $-=$
lower) the respective time arguments  $\tau_1$ and $\tau_2$ are found.
By a rotation $R$ in the Schwinger-Keldysh space defined by \cite{Keldysh} 
\begin{eqnarray}
R = \frac{1}{\sqrt{2}}
\left(
\begin{array}{lr}
{\color{white}-}1  &   {\color{white}-}1    \\
               -1  &   {\color{white}-}1
\end{array}
\right),
\end{eqnarray}
\begin{figure}[t]
\qquad \qquad\qquad\scalebox{1.18}[1.18]%
{\rotatebox{0}{\includegraphics[clip]{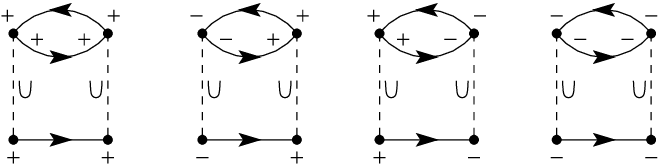}}}
\caption{Local self-energy $\Sigma^{\alpha\beta}$ 
  within the iterated perturbation theory (IPT) with its
  contributing four diagrams. The
  vertical dashed lines represent the interaction $U$. The solid lines
  correspond to the bath-Green's functions, the so-called Weiss-field,
  $\mathcal{G}^{\alpha\beta}$, cf. Eq. (\ref{Eq:DMFT}). 
  The IPT yields accurate results at half-filling \cite{Zhang}.
}
\label{Fig:IPT}
\end{figure}
the matrix Green's function can be rewritten in terms of the more familiar advanced and
retarded components of the Green's function according to
\begin{eqnarray}
G (\tau_1,\tau_2)
=
R^{-1}
\tilde{{\bf G} } (\tau_1,\tau_2)
R
=
\left(
\begin{array}{lcl}
0                     & \qquad &   G^{A}(\tau_1,\tau_2)   \\
G^{R}(\tau_1,\tau_2)   & \qquad &   G^{Keld}(\tau_1,\tau_2)
\end{array}
\right),
\label{Eq:Keldysh_Green-classics}
\end{eqnarray}
where $G^{Keld}(\tau_1,\tau_2)$ denotes the Keldysh component of the
Schwinger-Keldysh Green's function.
The assumed periodic driving of the atomic system encourages the use of the 
Floquet approach, see e.g. ref.  \cite{Haenggi}. Then the 
two-time Green's function  requires a generalized two-time Fourier transform
according to

\begin{eqnarray}
\nonumber
G^{\alpha\beta}_{m\, n\,;\, \sigma} ({\bf k}, \omega, \Omega_L)
&=&\!\!
\int_{-\infty}^{+\infty}\!\!\!\!\!\!\!\!\!\!
{\rm d}\tau_{rel} 
\frac{1}{\mathcal{T}} \int_{-\mathcal{T}/2}^{+\mathcal{T}/2}\!\!\!\!\!\!\!\!\!\!
{\rm d}\tau_{cm}
e^{i(\omega - \frac{m+n}{2}\Omega_L)\tau_{rel}}
\\ &&
\times e^{i(m-n)\Omega_L\tau_{cm}}
 G^{\alpha\beta}_{\sigma}({\bf k}, \tau_{rel}, \tau_{cm}),
\end{eqnarray}
 where $m,n$ are the Floquet indices labeling the Floquet modes of the
system, which are interpreted as the quantized lattice oscillations,
the phonons. The system is constrained to absorb and emit energy
in multiples of energy quanta $\hbar\Omega_L$. 
The Keldysh indices are $\alpha,\beta=\pm$, indicating the 
branch of the Keldysh contour, and 
$\mathcal{T}=\frac{2\pi}{\Omega_L}$ is the
time period. The system-time is shifted to a center-of-motion time
$\tau_{cm}=\frac{\tau_1 + \tau_2}{2}$ and a 
relative time coordinate $\tau_{rel}=\tau_1 - \tau_2$.
For completeness and later comparison, we note that for the non-interacting
case, {\em i.e.}, $U/D=0$, the Hamiltonian Eq. (\ref{Hamilton1}) can be solved
analytically, yielding 
\begin{eqnarray}
\label{non-interacting-G}
G_{mn}^{R}({\bf k}, \omega, \Omega_L) 
=
\sum_{\rho = -\infty}^{\infty}
\frac
{
J_{\rho-m}\!\!\left( \frac{E}{\hbar\Omega_L}
+
T\frac{\tilde{\epsilon}_{\bf k}}{\hbar\Omega_L} \right)
\!\!J_{\rho-n}\!\!\left( \frac{E}{\hbar\Omega_L}
+
T\frac{\tilde{\epsilon}_{\bf k}}{\hbar\Omega_L} \right)
}
{
\hbar\omega -\rho\hbar\Omega_L - \epsilon_{\bf k} + i \mathcal{O} 
},
\end{eqnarray}
where we summed over the spins $\sigma$, since they are not influenced by the
modulation of the potential.
In the above Eq. (\ref{non-interacting-G}) we introduced Bessel functions
$J_{\rho-m}$.
We note that $\epsilon_{\bf k}$ is the dispersion induced by the standard
hopping $t$, i.e. its Fourier transform. 
Furthermore $\tilde{\epsilon}_{\bf k}$ originates in the same way
from the modulation-induced 
hopping contribution $T(\tau)$ and is therefore time dependent.
Apart from the definition of $\epsilon_{\bf k}$, we
exclude $T$ from the definition $\tilde{\epsilon}_{\bf k}$.
Finally,  $\rho$ is the integer summation index.

To solve the full, {\em i.e.}, driven and interacting system ($U \ne 0$) at
zero temperature and half filling, we generalize a 
dynamical mean field theory (DMFT) to non-equilibrium. 
The DMFT \cite{Georges, Monien} maps the interacting lattice system onto 
a local impurity model embedded in a bath, which consists of all remaining
lattice sites in integrated form.
The local impurity  described by a local lattice Green's function 
$G_{mn}^{\alpha\beta}(\omega)$ is
related to the bath Green's function $\mathcal{G} (\omega)$, 
the so-called Weiss-field, by the DMFT self-consistency equation. 
The local self-energy appearing in the local
lattice Green's function depends on the Weiss-field, thus closing the
self-consistency. The calculation of the self-energy requires further
assumption and is achieved by invoking a so-called impurity solver,  the 
iterated perturbation theory (IPT)  \cite{Zhang}, which is here also generalized
to non-equilibrium. The IPT, a diagrammatic method, is demonstrated 
in Fig. \ref{Fig:IPT}.
The DMFT self-consistency equation for the Hamiltonian Eq. (\ref{Hamilton1}) 
in the above introduced Schwinger Keldysh Floquet space is derived as
\begin{eqnarray}
\label{Eq:DMFT}
\!\!\!\!\!\!&&\!\!\!\!\!\!\!\!\!\!\!\!\!\!\!\!\!\!\!\!\!\!\!\!\!\!\!\!\!\!\!\!\!\!\!\!\!%
\left[\mathcal{G}^{-1} (\omega)\right]_{mn}^{\alpha\beta}
= \left[ g_0^{-1} (\omega) \right]_{mn}^{\alpha\beta}    -   
\alpha\beta\,\,\,t\,  G_{mn}^{\alpha\beta}(\omega)\,\,\,t \\ 
\!\!\!\!\!\!\!\!\!&+&\alpha \delta_{\alpha\beta} 
\frac{E}{2}
\left[
\delta_{m,n+1}
+
\delta_{m,n-1}
+
\delta_{m+1,n}
+
\delta_{m-1,n}
\right] \nonumber\\ 
\!\!\!\!\!\!\!\!\!&+&\alpha\beta \left[\frac{T}{2}
\left(  
G_{m-1,n}^{\alpha\beta}(\omega)
+
G_{m+1,n}^{\alpha\beta}(\omega)
\right)t
-t\left(  
G_{m,n+1}^{\alpha\beta}(\omega)
+
G_{m,n-1}^{\alpha\beta}(\omega)
\right)\frac{T}{2} \right]  \nonumber\\ 
\!\!\!\!\!\!\!\!\!&+&\alpha\beta\left[\frac{T}{2}
\left( 
G_{m+1,n+1}^{\alpha\beta}(\omega)
+
G_{m-1,n-1}^{\alpha\beta}(\omega)
+
G_{m+1,n-1}^{\alpha\beta}(\omega)
+
G_{m-1,n+1}^{\alpha\beta}(\omega)
\right)\frac{T}{2}\right]. \nonumber
\end{eqnarray}
In the above equation, Eq. (\ref{Eq:DMFT}), the first line on the r.h.s. 
compares directly to the equilibrium expression, where the last term in the
first line marks the hopping $t$ on to a single site in the lattice 
 and off this single site (often called impurity). 
Consequently, $\left[ g_0^{-1} (\omega) \right]_{mn}^{\alpha\beta}  =
\alpha\delta_{\alpha\beta} (\omega -n\Omega_L )\delta_{nm}$. 
The second line represents the contribution of the onsite energy modulation 
originating from the first term on the r.h.s. of the Hamiltonian 
Eq. (\ref{Hamilton1}), the Kronecker delta symbols have to be interpreted as
the different absorption and emission processes of lattice quanta which 
contribute here. The remaining lines of Eq. (\ref{Eq:DMFT}), however, 
represent the part originating from hopping modulations in the driven system.
For instance in the third line, processes are found to be characterized by a standard
kinetic hopping $t$ on (off) the impurity combinded with a phonon induced 
hopping $T$ off (on) the impurity. 
The last line has  to be interpreted as the dynamics where both the hopping on
and  off the impurity are phonon induced.  In Eq. (\ref{Eq:DMFT}) products of the form
$\alpha\beta$ assume either the value $+1$ if $\alpha=\beta$ or $-1$ otherwise. 
The DMFT, Eq. (\ref{Eq:DMFT}) in conjunction with Fig. \ref{Fig:IPT},
offers, therefore, a solution for a matrix Green's function, which is
of matrix dimension $2 \times 2$ in Schwinger Keldysh space, cf. 
Eq. (\ref{Eq:Keldysh_Green-classics}), and of 
matrix dimension $n \times n$ in Floquet space. The index $n$ marks the number
of involved Floquet side bands in the problem.

\begin{figure}
 \qquad\qquad\qquad\scalebox{0.45}[0.45]{\rotatebox{0}{\includegraphics[clip]{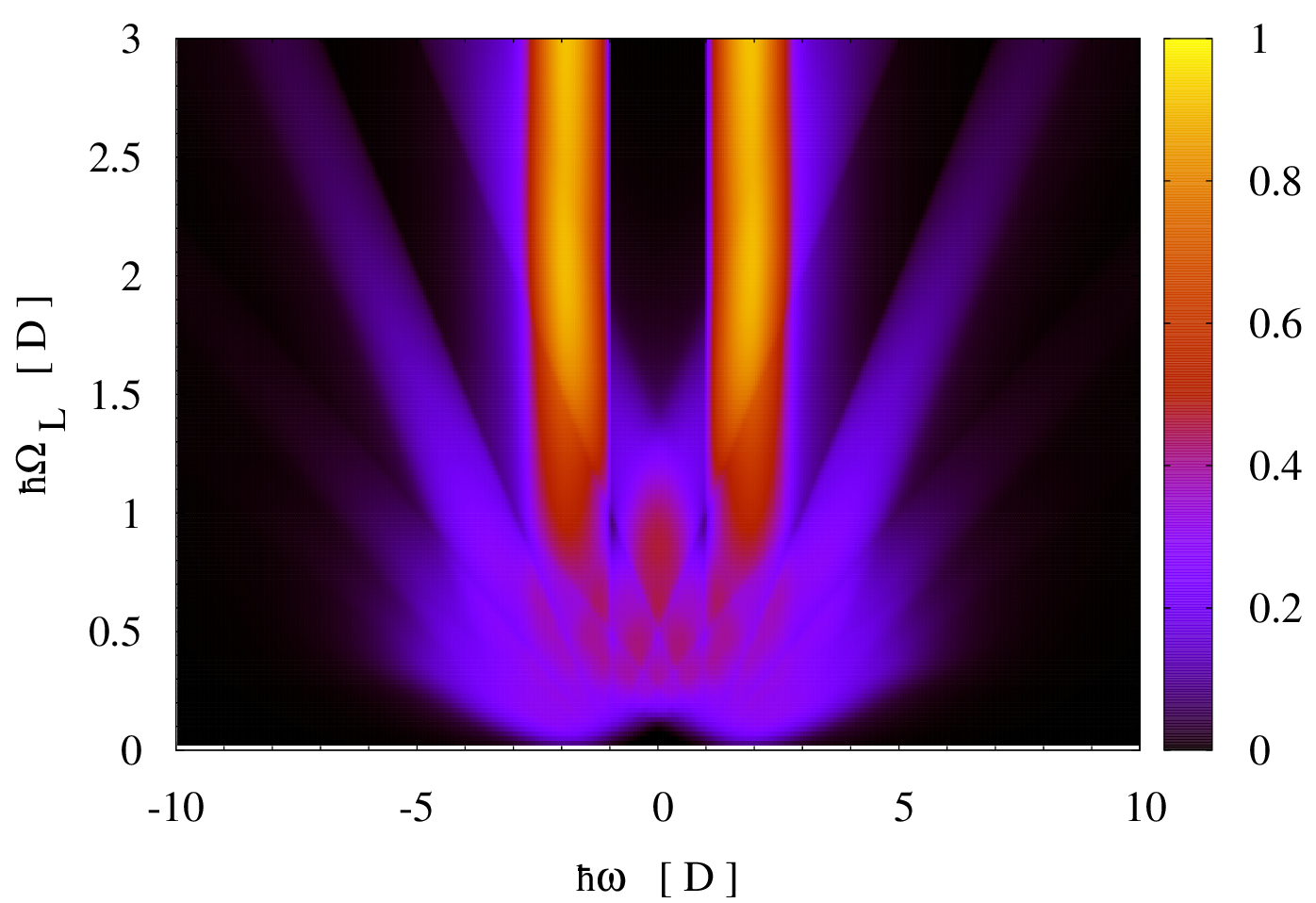}}}\\
\caption{(Color online) The local density of states (LDOS)
  $N(\omega,\Omega_L)$, Eq. (\ref{Eq:DoS}), as a function of atomic energy
  $\hbar\omega$ and modulation energy $\hbar\Omega_L$. 
$T/D=3.0$,  $U/D=4.0$ and $E/D=1.0$. The LDOS is displayed as a function of
atomic energies $\hbar\omega$ and lattice modulation frequencies $\hbar\Omega_L$.
For the limit of $\hbar\Omega_L \rightarrow 0$ we find that 
$N(\omega,\Omega_L)$ is not directly comparable to the equilibrium ground state, but
instead a new ground state is reached which features the AOC (see text).
}
\label{plot_f1}
\end{figure}

\begin{figure}
 \qquad\qquad\qquad\scalebox{0.39}[0.39]{\rotatebox{0}{\includegraphics[clip]{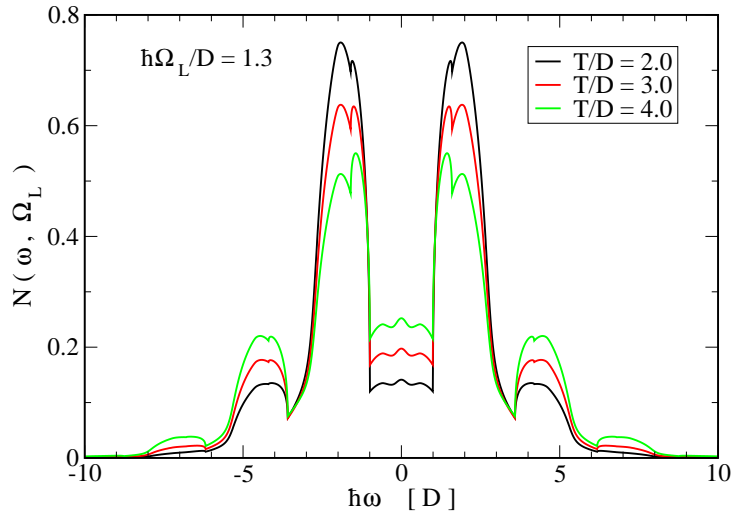}}}
\caption{(Color online) 
  $N(\omega,\Omega_L)$ as a function of atomic energy
  $\hbar\omega$ and for a single  modulation energy $\hbar\Omega_L / D =1.3$.
Displayed are three different hopping strengths $T/D=2.0;3.0;4.0$, other
parameters as in Fig. \ref {plot_f1}. With increasing hopping the gap states
are created and further features are weakened.
}
\label{plot_f1schnitte}
\end{figure}

\section{Results}

This substantial numerical effort results in
the full knowledge of the non-equilibrium Floquet-Keldysh-Green's function,
revealing  {\em e.g.}, the local density of states (LDOS), the non-equilibrium
distribution function, and the relaxation times by means of the self-energy.
The assumed initial state is the groundstate of the equilibrium system.
From the numerically computed components of the Green's function, 
we define the local density of states $N(\omega,\Omega_L)$ by 
the following expression,
where  momentum is integrated out and 
Floquet indices are summed
\begin{eqnarray}
N(\omega,\Omega_L)
=
-\frac{1}{\pi}\sum_{mn} \int {\rm d^3}k  {\rm Im\,} 
G^{R}_{mn} ({\bf k},\omega, \Omega_L).
\label{Eq:DoS}
\end{eqnarray}
We define the total non-equilibrium distribution function
$F^{neq}(\omega,\Omega_L)$
by the relation
\begin{eqnarray}
\sum_m G_{0m}^{Keld} (\omega,\Omega_L)
=
-2\pi i
\left[
1 \!-\! 2 F^{neq}(\omega,\Omega_L)
\right]
\frac 1 {\pi}
\sum_n {\rm Im} G_{0n}^{A}  (\omega,\Omega_L),
\end{eqnarray}
resulting in the definition of the total distribution function as
\begin{eqnarray}
F^{neq}(\omega,\Omega_L)
&=&
\frac {1}{2}
\left(
1
+
\frac{1}{2i}
\frac{\sum_m G_{0m}^{Keld}(\omega,\Omega_L)}{\sum_n{ \rm Im}G_{0n}^{A}(\omega,\Omega_L)}
\right).
\label{Eq:F_neq}
\end{eqnarray}
\begin{figure}
   \qquad\qquad\qquad\scalebox{0.45}[0.45]{\includegraphics[clip]{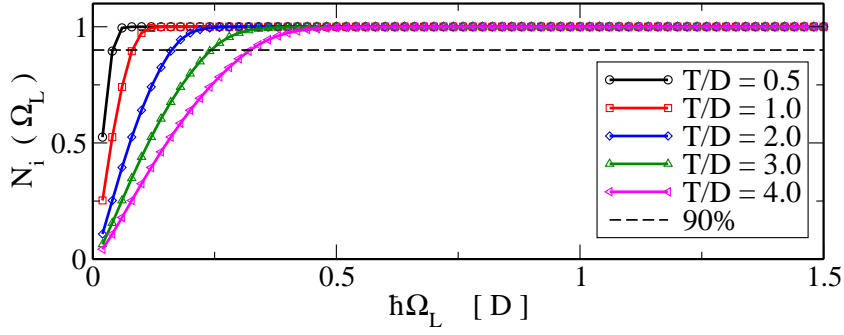}}
\caption{(Color online)
Atomic frequency integrated density of states $N_i(\Omega_L)$, 
see Eq. (\ref{Eq:N_integrated}), which is normalized to 1.
Numerical evaluation for various induced hopping strengths $T/D$ as shown, 
onsite repulsion strength is $U/D=4.0$ and $E/D=1$. The deviations from the value
1 are a measure for the range of validity of the numerics. The dashed
horizontal line marks 90\% of the norm result,  $N_i(\Omega_L)$ above the line
indicates a numerical error of less than 10\%, considered as  valid result. 
}
\label{fig_03}
\end{figure}

A  solution for the LDOS, Eq. (\ref{Eq:DoS}),
is shown in Fig. \ref{plot_f1}. There the development of pronounced Floquet side bands in
the LDOS structure is discussed.  
Especially in Fig. \ref{plot_f1schnitte} the LDOS for single external frequency but for three different hopping amplitude
$T$ is shown.
Distinct gap states evolve which induce a transition from the Mott insulator state
to the liquid phase. Both features result in severe changes of the fermionic
band structure and therefore cause significant changes of, {\em e.g.}, optical and
conduction properties.  
The behavior of the density of states as a
function of the external modulation energy $\hbar\Omega_L$ exhibits two
limiting cases with a cross-over regime in between them.
For the limit of small modulation frequencies $\hbar\Omega_L \rightarrow 0$ it is
interesting to note that
all Floquet modes gain more and more spectral weight (compare also Fig. \ref{fig_03}).
This signals the onset of an orthogonality catastrophe,  as predicted by P. W.
Anderson (AOC) \cite{AOC}.
Anderson states that the ground state of the system is the equilibrium
state whereas the zero quasiparticle state does not exist in the fermionic
system. The limit of that zero quasiparticle state would correspond to an
infinite number of contributing Floquet modes and that 
state would mark a new ground state which is orthogonal to
the original equilibrium state, caused by the change in the potential of the
system shown in Fig. \ref{plot_f1}. 
Technically, this is seen as a drastic enhancement of the arguments of the
Bessel functions, {\em e.g.},  for the 
non-interacting expression in Eq. (\ref{non-interacting-G}).

At this point, it should be emphasized that any 
numerical evaluation is always limited to treatments 
with a finite number of Floquet modes. 
Therefore the utilized numerical implementation 
is optimized towards the
limit of maximum validity at a minimal cost and affordable amount of
computational effort. 
An analysis of the numerical validity in terms of the normalized and
frequency integrated density of states 
\begin{figure}
\qquad \qquad\qquad\scalebox{0.45}[0.45]{\rotatebox{0}{\includegraphics[clip]{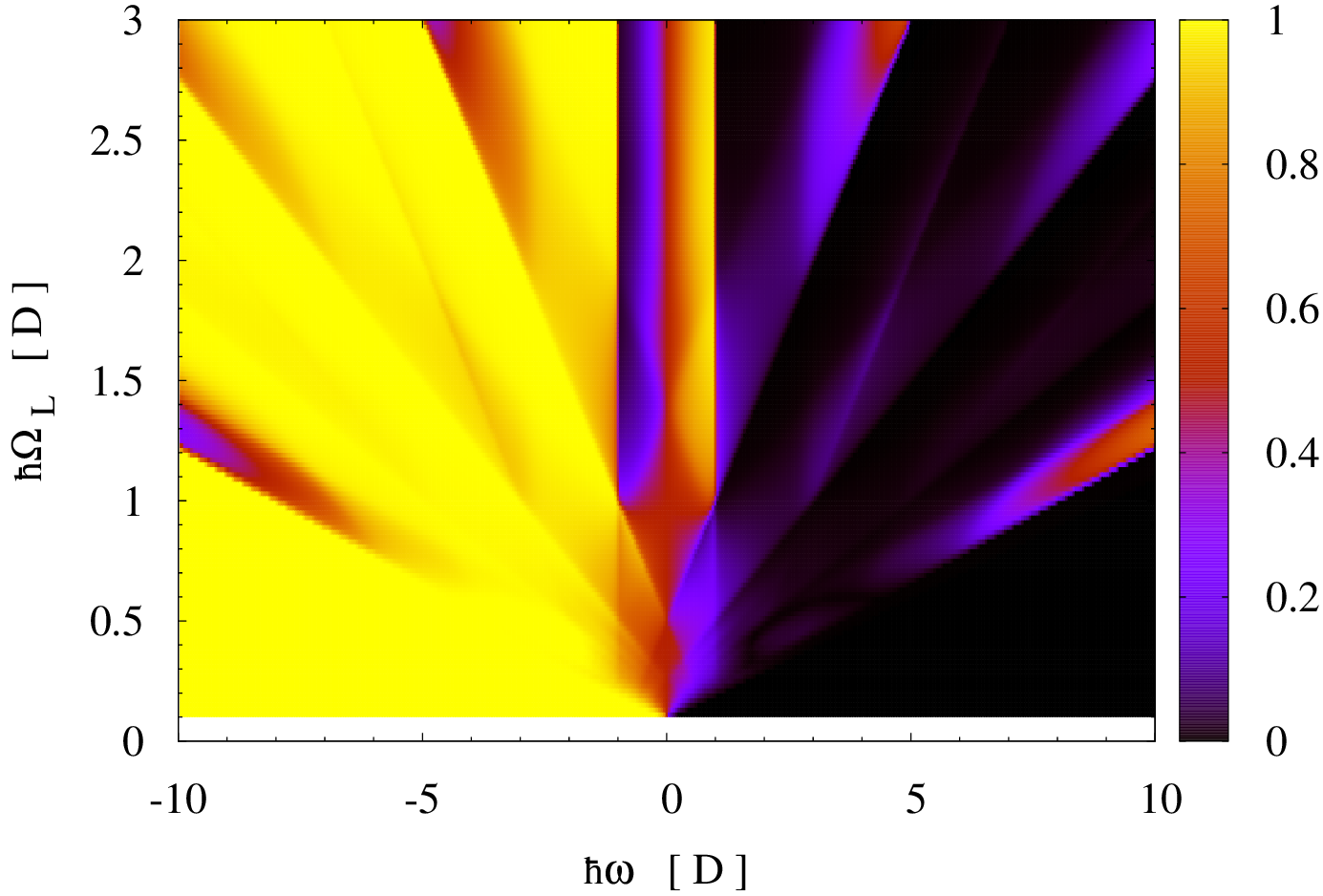}}}
\caption{(Color online)
Displayed is the non-equilibrium distribution function 
$F^{neq}(\omega,\Omega_L)$ in Eq. (\ref{Eq:F_neq}),
as a function of atomic energy $\hbar\omega$ and
external modulation energy $\hbar\Omega_L$.
The  parameters for this original insulating state are $U/D=4.0,$ $T/D=2.0,$
$E/D=1.0$. Note the behavior around external modulation energies of
$\hbar\Omega_L/D \simeq 1$ in the gap region $-1<\hbar\omega/D<1$ (see also text).
}
\label{fig_05}
\end{figure}
\begin{eqnarray}
N_{i}(\Omega_L)
:=
\int{\rm d}\omega N(\omega,\Omega_L)=1
\label{Eq:N_integrated}
\end{eqnarray}
 can be found in Fig. \ref{fig_03}.
As a function of the lattice oscillation frequency $\hbar\Omega_L$, 
the numerical value of the LDOS
displays, deviations from its normalization constant $1$. 
The larger the deviation, the larger the encountered numerical error. 
In   Fig. \ref{fig_03}, $N_i(\Omega_L)$ above the horizontal dashed line 
indicates  results with a numerical error of less than $10\%$, which
we will consider valid results here. For increasing hopping strengths $T$, the
numerical accuracy is succeedingly decreased 
(all other parameters remain unchanged)
for small modulation frequencies $\Omega_L$. The
physical interpretation of the infinite increase of the number of
Floquet-Keldysh-Green's modes contributing for small modulations $\Omega_L$
indicate the AOC, see above.
Even for the largest considered $T$, numerical results for 
modulation energies  $\hbar\Omega_L/D > 0.25$ can be 
considered as accurate within an error range of less than 
10 \%  for the used implementation.
 
\begin{figure}
\qquad\qquad\qquad\scalebox{0.6}[0.6]{\rotatebox{0}{\includegraphics[clip]{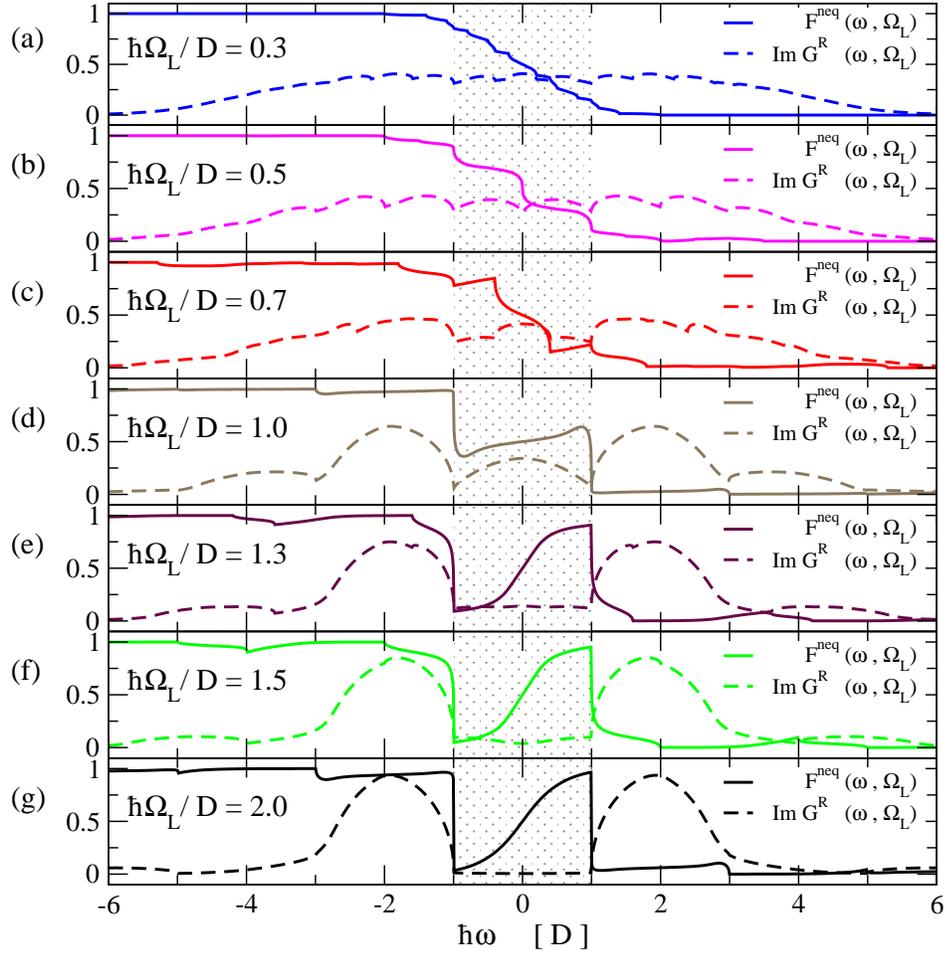}}}
\caption{(Color online)
Displayed is $F^{neq}(\omega,\Omega_L)$, Eq. (\ref{Eq:F_neq}),
together with the corresponding 
LDOS $N(\omega,\Omega_L)$, Eq. (\ref{Eq:DoS}),
for parameters
$U/D=4.0$ , $T/D=2.0$ $E/D=1.0$ at a series of different external modulation
frequencies , $\hbar\Omega_L/D\in \{0.3,0.5,0.7,1.0,1.3,1.5,2.0\}$, from
panel (a) to (g). The band gap in equilibrium has the width $2 D$, i.e. the
hatched region $-1<\hbar\omega/D<1$. We note that
with increased modulation energies trapping effects of fermions in the Mott
gap (hatched area) can be observed. The behavior of the derivative of the distribution
function with respect to the frequency $\omega$ at the Fermi edge severely
influences the conductivity (see Fig. \ref{fig_08}). 
}
\label{fig_05btest}
\end{figure}

In Fig. \ref{fig_05btest} we discuss the behavior of the LDOS,
Eq. (\ref{Eq:DoS}) 
and the occupation number for increasing external 
modulation energy $\hbar\Omega_L$. 
For frequencies $\hbar\Omega_L/D < 1$ the behavior of the ultracold
Fermi gas changes from Mott insulating to liquid or conducting. Pronounced
Floquet side bands
\cite{Haenggi} develop and intersect in between the Hubbard bands. The Mott gap
almost disappears and a liquid or conducting 
regime is established, where the liquid density of
states can be continuously driven by the external modulation. 
For the occupation number of the gap states  a step-like behavior for long
wavelength modulations is found 
 which can be interpreted as the absorption or emission
of energy quanta (phonons). In the non-equilibrium fermionic distribution
function we derive that for long wavelength modulations the majority of
fermions resides in states below the Fermi edge ($\hbar\omega=0$).

The two limiting regimes of small and large lattice modulations are separated
by a cross-over at $\hbar\Omega_L/D \simeq 1$ (see Fig. \ref{plot_f1}). 
At the crossing, the modulation-induced Floquet side bands are 
forced to intersect (thus crossing) in the gap and acquire a
maximum of spectral weight (see Fig. \ref{fig_05}) in this area. 
Moreover the occupancy
from the states right above the lower Hubbard band is shifted towards states
right below the upper Hubbard band and additionally the entire gap is almost
equally occupied. We further remark that the
excitation behavior to reach the upper Hubbard band at
the crossing changes from virtual, {\em i.e.}, successive absorption, to direct.  

Right above the crossing, for external modulation frequencies $\hbar\Omega_L/D > 1$, we
find that the liquid behavior is dramatically reduced.  In the lower panels of 
Fig. \ref{fig_05btest} we discuss the LDOS and the distribution 
function for external frequencies above $\hbar\Omega_L/D=1$. 
Note here 
that above the crossing the LDOS shows a significant change in the
gap. The
step-like structure vanishes and  exhibits for further
increasing modulation energies $\hbar\Omega_L$ almost 
the behavior of a Mott insulator 
with an unconventional
occupation number in the Mott gap in conjunction with a weak spectral weight
there. 
Combining both, the discussion concerning the spectral weight and the
distribution number right above $\hbar\Omega_L/D =1$, we find a pronounced shift of
occupation which resembles a trapping of particles right below the upper
Hubbard band, which is an inversion-like situation. This means, 
atoms occupy energy states above the Fermi energy, especially in the gap
region, therefore establishing a population inversion as found and used in
other systems to start and maintain lasing behavior.
The relaxation of the
Fermions is impossible because no unoccupied states 
are within reach 
for emission processes of an integer number of phonons of the periodic
modulation. This effect establishes the atomic population inversion in the
pumped system and requests  experimental verification. A utilization of the
population inversion for other experimental or technological methods might be
rather promising, e.g. for phonon pumped lasing.
For significantly 
faster lattice vibrations of the confining potential the system is
not able to follow the perturbations and returns to a stationary state similar
to equilibrium. The trapping of the occupation is however preserved.

\begin{figure}
\qquad\qquad\qquad\scalebox{0.45}[0.45]{\rotatebox{0}{\includegraphics[clip]{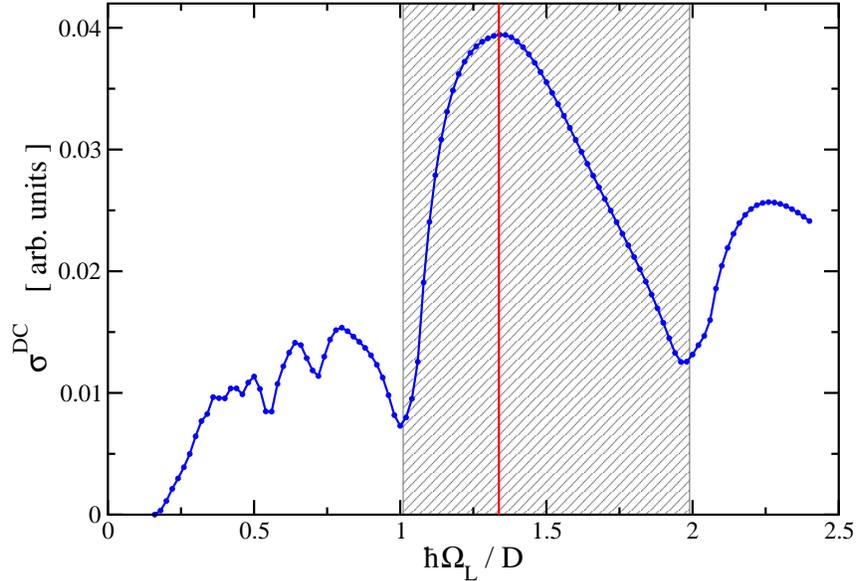}}}
\caption{(Color online)
DC conductivity $\sigma^{{\rm   DC}}(\Omega_L)$, Eq. (\ref{sigma_dc}), as a function of external 
modulation energy $\hbar\Omega_L$.
The highest value $\sigma^{DC}$ corresponds to an external frequency of 
$\hbar \Omega_L / D \simeq 1.3$ (vertical line). 
The conductivity not only depends on the absolute value of the LDOS but
also on the sign of the derivative of the distribution function
$F^{neq}(\omega,\Omega_L)$. Parameters are the same as in
Fig. \ref{fig_05btest}.
The hatched area marks the energy region where  two-phonon processes excite
atoms from the lower to the upper Hubbard band.
}
\label{fig_08}
\end{figure}

The closing of the Hubbard gap within an intermediate range of the external
modulation frequency is also observed by 
the calculated DC conductivity, which can be written in the form

\begin{eqnarray}
\label{sigma_dc}
\sigma^{{\rm DC}}(\Omega_L)
&\!\!=\!\!&
\lim_{\omega \rightarrow 0} \sum_{m} {\rm Re} \,\, \sigma_{0m}   (\omega,\Omega_L)
 \\
&\!\!=\!\!&
\sum_{m} 
\frac{8e^2t^2}{2\pi^3}\!\!
\int\!\! {\rm d}{\epsilon}
N_0(\epsilon)
\!\!\int\!\! {\rm d}{\omega \,'}\!
\left({\rm Im\,}G^{R}_{0m}
(\epsilon, \omega \,',\Omega_L)\right)^2 \!\!
\frac{\partial}{\partial \omega\,'}  F^{neq}_{0m}(\omega\,', \Omega_L),
\nonumber
\end{eqnarray}
where $e$ is the elementary charge, $t$ the hopping amplitude
and $N_0(\epsilon)$ is the bare density of states.
The nonequilibrium distribution $F^{neq}_{0m}(\omega,\Omega_L)$
is defined by the relation
\begin{eqnarray}
G_{0m}^{Keld} (\omega, \Omega_L)
&=&
-2\pi i
\left[
1 - 2 F^{neq}_{0m}(\omega, \Omega_L)
\right]
\frac 1 {\pi}
{\rm Im} G_{0m}^{A}  (\omega,\Omega_L)\\
F^{neq}_{0m}(\omega,\Omega_L)
&=&
\frac {1}{2}
\left(
1+\frac{1}{2i}\frac{G_{0m}^{Keld} (\omega, \Omega_L)}
{{\rm Im} G_{0m}^{A}  (\omega, \Omega_L)}
\right),
\end{eqnarray}
where the Keldysh and the advanced component of the Green's function
result from the numeric DMFT solution.

The numerical evaluation of the above  DC conductivity 
$\sigma^{{\rm   DC}}(\Omega_L)$, Eq. (\ref{sigma_dc}),
is presented in Fig. \ref{fig_08}. The parameters are 
the same as those given in the caption of 
Figs. \ref{fig_03} and \ref{fig_05}. 
We find a strong dependency of $\sigma^{{\rm   DC}}(\Omega_L)$ 
with respect to the sign changes of the
distribution function $F^{neq}_{0m}(\omega, \Omega_L)$. 
The distinct global maximum in the
range  $1\le \hbar\Omega_L /D\le 1.75$ accounts for the intermediate regime, where
the absolute height is dominated by the value of the LDOS at the Fermi edge. 
The somewhat less pronounced peak at larger frequencies $\hbar\Omega_L/D>2$, 
however, is also attributed to the strong population inversion in 
the regime of external lattice modulations.
The dominant and lowest in energy process is a two phonon process. Given the
width of the gap $\Delta/D = 2$ for $U/D=4$, an excitation energy of $\hbar\Omega_L/D=1$ 
will suffice to bridge the gap and to transfer fermions to the upper Hubbard band. 
The conductivity in Fig. \ref{fig_08} also exhibits this behavior. The maximum of 
$\sigma^{{\rm   DC}}(\Omega_L)$ between $ 1 < \hbar\Omega_L/D < 2 $ reflects just this
excitation behavior. Two phonons, each with energy $\hbar\Omega_L/D=1$, are absorbed 
by one fermionic atom, therefore the atomic energy is increased by
$\hbar\omega/D=2$, the amount of energy one atom resting at the upper edge of the lower Hubbard band needs
to be pumped to a state just above the lower edge of the upper Hubbard band.
Consequently, a fermion absorbing two phonons each with  energy of $\hbar\Omega_L/D=2$  
raises the atomic energy by $ \hbar\omega / D = 4 $. A fermion at the lower edge of the 
lower Hubbard band gaining this amount in energy is transferred to just below the upper edge 
of the  upper Hubbard band. Finally, in Fig. \ref{fig_08} the form of 
the conductivity Eq. (\ref{sigma_dc})
outside the interval $ 1 < \hbar\Omega_L/D < 2 $  is caused by higher Floquet bands 
involving more than two phonon processes or by excitation dynamics between individual 
Floquet bands instead of in between the Hubbard bands.
The dominant contribution, however, is the lowest excitation in 
between the two equilibrium Hubbard bands.

\section{Conclusion}

A theory of ultracold fermionic
atoms described by a Hubbard model including strong repulsive interactions is
discussed. The quantum
criticallity is derived  with periodic lattice potential modulations which
drive the considered system out of thermodynamical equilibrium.
By investigating a Floquet-Keldysh-Green's function approach we find a
cross-over at zero temperature between
two limiting characteristics, the AOC for $\hbar\Omega_L/D \rightarrow 0$ and a
quasi-equilibrium solution for $\hbar\Omega_L/D\rightarrow \infty$. 
Pronounced side-bands lead to a rather complicated density of states in
the gap for $\hbar\Omega_L/D < 1$, which indicates a transition to the
liquid or conducting regime right at the onset of the modulation. In the
vicinity of the cross-over we find a maximum of spectral weight inside the
original excitation gap. For external frequencies $\hbar\Omega_L/D>1$,  population
trapping in the gap is observed. Beyond, the system approaches an
equilibrium-like Mott insulator regime, which indicates that the 
system is not able to follow fast perturbations.

\section{Acknowledgments}

The author thanks A. Lubatsch, H. Monien and G. Sch\"on for stimulating and
fruitful discussions. Special thanks go to V. Dittrich for reading this manuscript
critically. Support by Karlsruhe School of
Optics and Photonics (KSOP) is acknowledged. The author is fellow of the Athene
program funded by the excellence initiative of the federal government of Germany.\\

\end{document}